\documentclass[aps,prl,reprint,twoside,floatfix,superscriptaddress,nolongbibliography,letterpaper]{revtex4-1} %as an APL paper

\usepackage{graphicx}
\usepackage{amsmath}
\usepackage{amssymb}
\usepackage{amsfonts}
\usepackage{dcolumn}
\usepackage{epsfig}
\usepackage{bm}

\begin{document}

\title{Frequency Conversion in a High $Q$-factor Sapphire Whispering Gallery Mode Resonator due to Paramagnetic Nonlinearity}

\author{Creedon, D.L.}
\email{daniel.creedon@uwa.edu.au}
\affiliation{ARC Centre of Excellence for Engineered Quantum Systems, University of Western Australia, 35 Stirling Highway, Crawley WA 6009, Australia}

\author{Benmessa\"i, K.}
\affiliation{ARC Centre of Excellence for Engineered Quantum Systems, University of Western Australia, 35 Stirling Highway, Crawley WA 6009, Australia}

\author{Tobar, M.E.}
\affiliation{ARC Centre of Excellence for Engineered Quantum Systems, University of Western Australia, 35 Stirling Highway, Crawley WA 6009, Australia}

\date{\today}

%%%%%%%%%%%%%%%%%%%%%%%%%%%%%%%%%%%%%%%%%%%%%%%%%%%%

\begin{abstract}
Nonlinear frequency conversion is a well known and widely exploited family of effects in optics, often arising from a Kerr nonlinearity in a crystal medium. Here, we report high stability frequency conversion in the microwave regime due to a $\chi^{(3)}$ nonlinearity in sapphire introduced by a dilute concentration of paramagnetic spins. First, we produce a high stability comb from two microwave fields at 12.029 and 12.037 GHz corresponding to two high $Q$-factor Whispering Gallery (WG) modes within the Electron Spin Resonance (ESR) bandwidth of the Fe$^{3+}$ ion. The resulting comb is generated by a cascaded four-wave mixing effect with a 7.7 MHz repetition rate. Then, by suppressing four-wave mixing by increasing the threshold power, third harmonic generation is achieved in a variety of WG modes coupled to various species of paramagnetic ion within the sapphire.
\end{abstract}

\maketitle
Historically, high $Q$-factor whispering gallery (WG) mode resonators have been used as frequency determining elements in high stability, low-noise microwave oscillators, both at room temperature and cryogenic temperatures\cite{Locke2006rsi,Tobar95}. More recently, a cryogenic whispering gallery maser oscillator was developed\cite{Pyb2005apl,Benmessai2007el,Benmessai2008prl,BenmessaiAmpProc,BenmessaiGyrotropic,Creedon2010,BenmessaiPRB2012} in which the sapphire resonator acts as the gain medium for maser oscillation due to a parts-per-billion concentration of Fe$^{3+}$ impurities in the sapphire lattice. Annealing in air has led to mass conversion of Fe$^{2+}$ ions in the lattice to Fe$^{3+}$, resulting in a 3 order of magnitude increase in active ion concentration, up to 150 ppb \cite{Creedon2010}. At this level of concentration, nonlinear behaviour was recently observed due to an interaction between the paramagnetic impurity spins and high $Q$-factor ($Q\approx10^9$) WG modes near 12 GHz\cite{CreedonPRL2012}. This was modelled as a degenerate Four Wave Mixing (FWM) process due to a $\chi^{(3)}$ nonlinearity associated with the presence of paramagnetic Fe$^{3+}$ spins.  With only a single input field, an output spectrum of three signals was observed, which coincide with the pump, signal, and idler frequencies of the FWM scheme, all of which lie within a broad ESR bandwidth associated with the Fe$^{3+}$ spins. Weak frequency upconversion from the pump to the signal frequency was achieved by virtue of this spin induced nonlinearity.

In this work, we exploit the new found nonlinearity and investigate other mechanisms for frequency conversion in sapphire. By implementing a doubly-pumped scheme, which actively utilises two input fields, stable microwave comb generation is achieved with a spectrum of output signals measured across the whole ESR bandwidth of the Fe$^{3+}$ ions. Comb generation arising from Four Wave Mixing has been observed previously at optical frequencies in toroidal microcavities, exploiting the Kerr nonlinearity of the crystal medium\cite{KippenbergComb,DelHaye:2008uq,A.:2011kx}. Here, we extend this novel approach into the microwave regime by exploiting the $\chi^{(3)}$ nonlinearity arising from the population of paramagnetic spins in the resonator, rather than an intrinsic nonlinearity arising from the crystal structure. Impurity spins in solids have experienced significant research interest in recent years due to their potential application to quantum information storage and processing. To the authors' knowledge, our result is the first observation of pure microwave frequency comb generation due to a paramagnetic nonlinear process arising from such spins in a crystalline host. In addition to comb generation, suppressing the FWM effect by reducing the $Q$-factor of the 12.029 and 12.037 GHz modes (which in turn increases the threshold power for FWM), allowed us to measure a significant level of third harmonic generation\cite{stanciu:4064,APEX.2.122401} for frequencies corresponding to a wide variety of WG modes. No significant second harmonic generation\cite{2harmonic,Physics.3.32} is observed above what is generated spuriously in the measurement setup, which also confirms that the nonlinearity is third-order in nature.
\begin{figure}[b]
\includegraphics[width=2.8in]{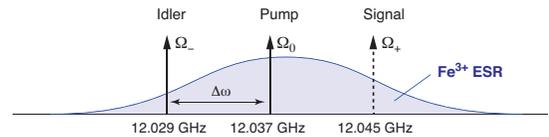}
\caption{\label{variables}(Not to scale.) Schematic of FWM within the inhomogeneously broadened Fe$^{3+}$ ESR.}
\end{figure}
\begin{figure*}[t]
\includegraphics[width=6in]{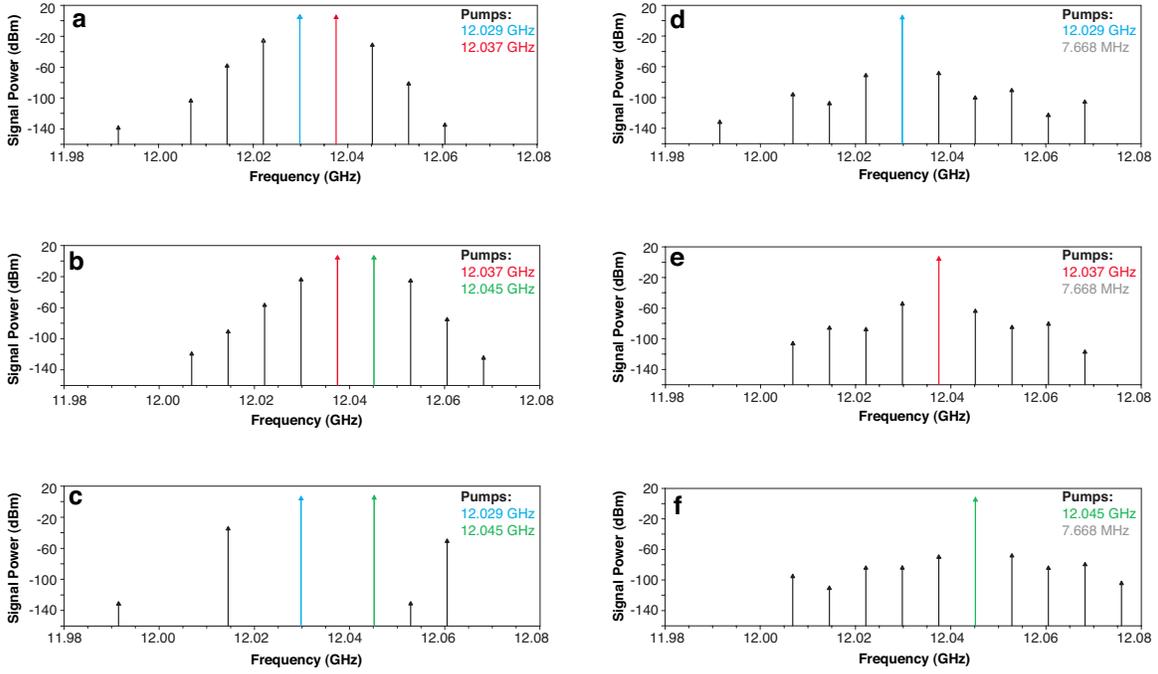}
\caption{\label{comb}Measured spectrum of signals as a result of a classical doubly pumped four-wave mixing scheme. (a) Crystal pumped at 12.029 GHz and 12.037 GHz, (b) 12.037 GHz and 12.045 GHz, (c) 12.029 GHz and 12.045 GHz, (d) 12.029 GHz and 7.668 MHz, (e) 12.037 GHz and 7.668 MHz, and (f) 12.045 GHz and 7.668 MHz.}
\end{figure*}
\begin{figure}[t]
\includegraphics[width=2.8in]{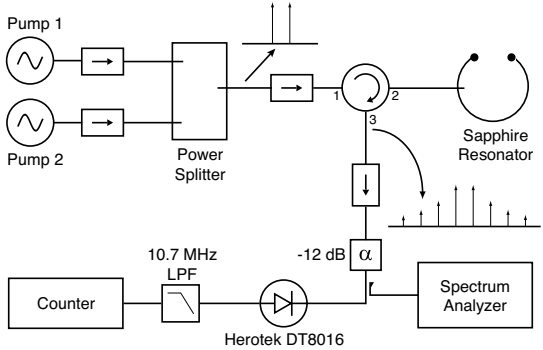}
\caption{\label{stabschematic}Schematic of the circuit used for comb generation and frequency stability measurement. The configuration of the resonator is the same as that used by Creedon et al. \cite{CreedonPRL2012}, however in the work described here, the resonator was examined in reflection using only the highest coupled port. The pump synthesizers and the counter were all referenced to a hydrogen maser.}
\end{figure}
\begin{figure}[t]
\includegraphics[width=2.8in]{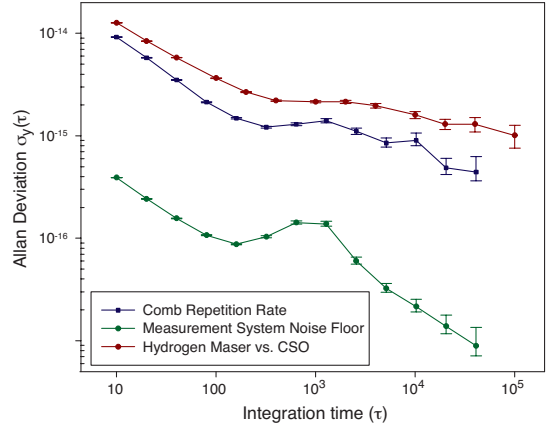}
\caption{\label{stab}Measured fractional frequency stability of the comb repetition rate, relative to the pump signal at 12.037 GHz. The  stability of the reference source used (Kvarz Hydrogen Maser) is shown, which was measured by comparison to a Cryogenic Sapphire Oscillator by Nand et al.\cite{nand}. The noise floor of the measurement system is shown, which was obtained by mixing the output of both synthesizers and passing the beat signal directly to the Agilent 53132A Frequency Counter.}
\end{figure}
\begin{figure}[t!]
\includegraphics[width=8.5cm]{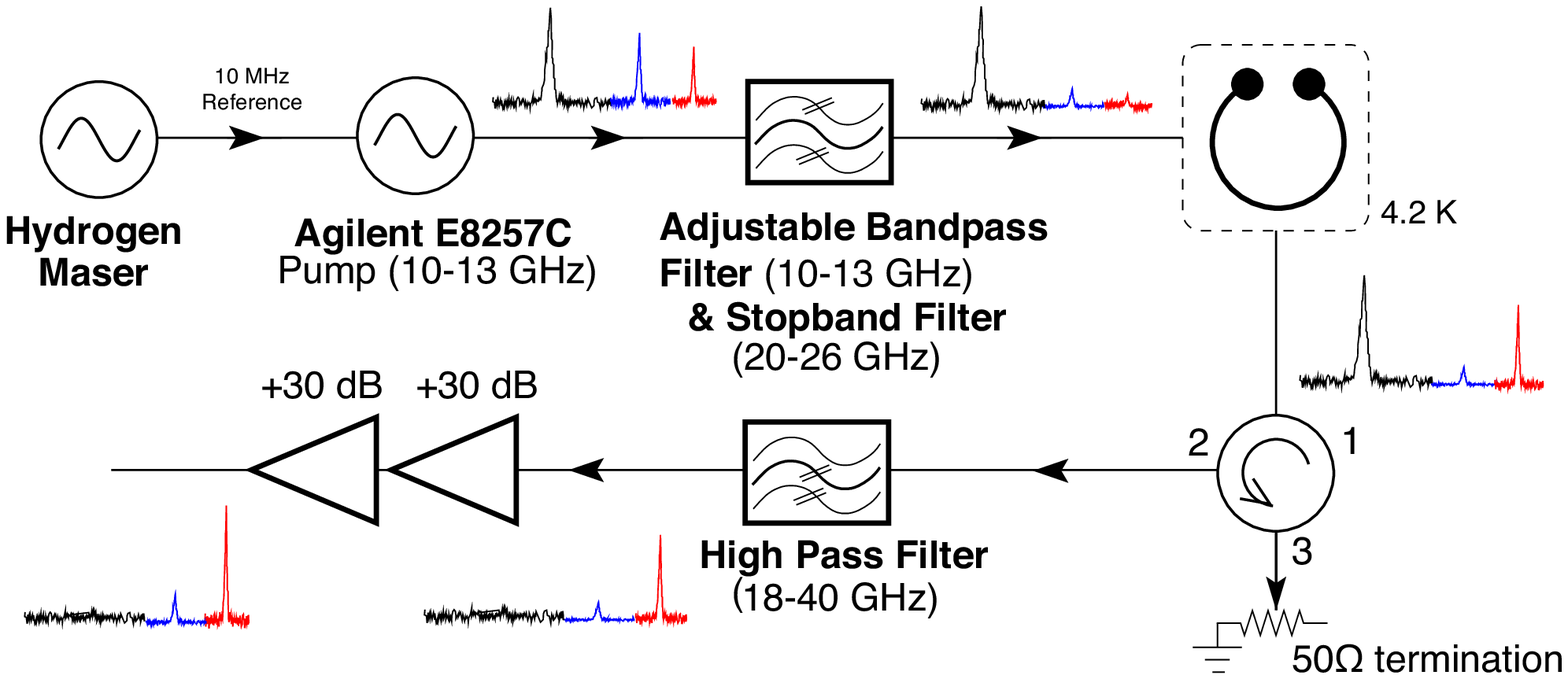}
\caption{\label{3rdharmonic} Schematic of the measurement system used to generate and detect the harmonics, as well as filter spurious harmonics generated by the synthesizer.}
\end{figure}
The experimental setup consists of a cylindrical sapphire monocrystal, 49.983 mm diameter $\times$ 30.002 mm height, which is cleaned in nitric acid in an ultrasonic bath. An axially protruding spindle at one end of the sapphire was used to mount it in a silver plated copper cavity. The resonator is machined such that the anisotropy $C$-axis of the sapphire is aligned with the cylindrical \textit{z}-axis, and two radially oriented loop probes couple microwave radiation in and out of the crystal. The loop probes were orthogonally oriented to reduce cross-coupling. The cavity was fastened under permanent vacuum in a small can, which was then mounted inside a vacuum chamber at the bottom of a long cryogenic insert placed into a 100 litre liquid helium dewar. A vector network analyser was used to examine the whispering gallery modes of interest in transmission and reflection, and to measure the probe couplings. Figure \ref{variables} shows a schematic model of the Fe$^{3+}$ spin system in the resonator in frequency space. $\Omega_0$ and $\Omega_{-}$ represent fixed microwave WG mode resonances in the sapphire resonator with a bandwidth of order 10 Hz. A broad Electron Spin Resonance, which encompasses both WG mode resonances in its 27 MHz half-bandwidth, exists centred around 12.04 GHz due to the presence of Fe$^{3+}$ ions in the sapphire lattice. The output signal of the FWM process, $\Omega_{+} = 2\Omega_0 - \Omega_{-}$, is shown dashed as no WG mode exists at 12.045 GHz.

Further to the results of Creedon et al.\cite{CreedonPRL2012}, we implemented a classical four-wave mixing scheme in the same system with the injection of two frequencies. This resulted in the generation of a microwave comb with a repetition rate at the difference frequency of the pump and idler, $\Delta\omega=$ 7.668 MHz, due to a cascaded effect in which each signal of the FWM acts as a source field for further mixing. Figure \ref{comb} shows the output of a comb of four-wave mixing signals, whose power effectively maps the ESR bandwidth of the Fe$^{3+}$ impurities. When pumped at 12.037 GHz and 12.029 GHz simultaneously (Fig. \ref{comb}(a)), the resultant signal at 12.045 GHz had a dramatically increased output power over the previous single-pump observation of FWM, where it was close to $-100$ dBm. It can be shown \cite{Ramirez2011} that the maximum quantum conversion efficiency of the degenerate four-wave mixing process which enables comb generation is given by
\begin{equation}
\eta_{max}=\frac{1}{2}\left(1+\frac{\Delta\omega}{\omega_{0}}\right)
\end{equation}
which we calculate to be $\sim$50\% for our dual-pump system. The significant increase in output power here indicates that the previous single-pump scheme, which did not generate a comb, was operating in a regime of low conversion efficiency. Of particular interest is the generation of four-wave mixing when pumped concurrently at 12.037 GHz and 12.045 GHz (Fig. \ref{comb}(b)), noting that no WG mode or cavity resonance exists at the latter frequency. We anticipate that in this case, the difference frequency of $\Delta\omega$ is generated within the crystal, which then allows the generation of photons at 12.029 GHz which are resonantly enhanced by a WG mode, after which efficient four-wave mixing can take place.  This was tested, and excitation of the comb is possible by pumping at any one of the $\omega_{-}$, $\omega_{0}$ or $\omega_{+}$ frequencies, at the same time as the $\Delta\omega$ frequency at 7.668 MHz, albeit with much lower conversion efficiency (Fig. \ref{comb}(d)-(f)).
\begin{figure}[b]
\includegraphics[width=3in]{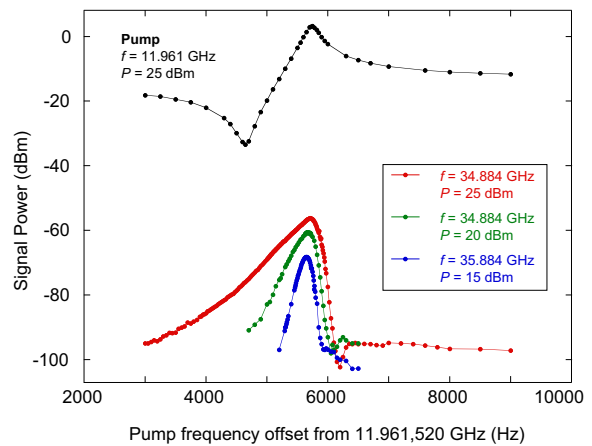}
\caption{\label{11961fig}Third harmonic generation at 35.884 GHz corresponding to a pump of 11.961 GHz exciting a WG mode resonance as shown in Figure \ref{3rdharmonic}}
\end{figure}
\begin{figure}[b]
\includegraphics[width=3in]{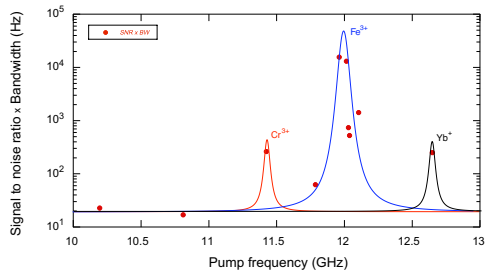}
\caption{\label{gbwfig}Product of $SNR\times BW$ at the third harmonic frequency as a function of pump frequency. This is a relative measure of the strength of the third harmonic generation and the $\chi^{(3)}$  nonlinearity, which is due to Cr$^{3+}$, Yb$^{+}$, and Fe$^{3+}$}
\end{figure}
The comb generated by the cascade of FWM processes is effectively band-pass filtered by the gain profile of the Fe$^{3+}$ ESR, which resonantly enhances the FWM signals. The final detectable comb contains approximately nine evenly spaced signals, depending on the input pump frequencies. To determine the frequency stability of the repetition rate of the comb ($\Delta\omega$), a beat of the output signals was measured as per Fig. \ref{stabschematic}. A microwave circulator was used to both inject the pump signals and read out the resultant comb through the most highly coupled port of the resonator cavity. A tunnel diode detector was used as a mixer to generate the beat frequency of the comb signals, which was then low pass filtered and measured by a high resolution frequency counter referenced to a Kvarz hydrogen maser. In our system, no comb stabilisation techniques were employed other than to control the temperature of the resonator at its frequency-temperature turnover point. The Allan Deviation of the comb beat was computed, and the instability relative to the excitation frequency was of order $2\times 10^{-15}$ at 100 seconds of integration, which exceeds the stability of the reference Hydrogen maser.

In addition to comb generation, the same nonlinearity introduced by the presence of Fe$^{3+}$ and other paramagnetic ions has also allowed the detection of significant third-harmonic generation. Figure \ref{3rdharmonic} shows the setup used to generate and measure these signals. The internal electronics of the microwave synthesizer in use also generate weak second and third harmonic tones, and thus great care is taken to filter these out and ensure a spectrally pure input signal to the resonator. In this way, we ensure that the detected harmonics are genuinely created through the nonlinearity of the crystal, rather than simple propagation of spurious harmonics originating in the synthesizer. To remove the synthesizer harmonics, the signal is first band pass filtered at the pump frequency set between 10 to 13 GHz, then passed through a tuneable stop band filter at which the second and third harmonics are attenuated considerably (60 dB between 20--26 GHz, and 30dB between 30--39 GHz). The pure pump signal is then injected into the resonator mounted at 4.2 K in a cryocooler, and the output signal is passed through a waveguide that attenuates frequencies outside the band lower than 18 GHz, resulting in 80 dB attenuation of the pump signal between 10 to 13 GHz. Finally, the 2$^{\text{nd}}$ and 3$^{\text{rd}}$ harmonics are amplified through a chain of two amplifiers with a total gain of 60 dB. The use of the waveguide is crucial to avoid any harmonic generation from the pump signal when passed through the amplifiers, which also act as nonlinear devices.

To characterise the second and third harmonic generation of the resonator, we measure the residual power generated by the system at these frequencies without the cavity, but with an equivalent attenuation included instead. Then by inserting the cavity we can define the signal-to-noise ratio, $SNR = P_{h}/P_{res}$, where $P_{res}$ is the residual harmonic power without the cavity and $P_{h}$ is the power measured at the generated harmonic frequency with the cavity in place. Figure \ref{11961fig}  shows the value of $P_{h}$ for the third harmonic at 35.884 GHz over a variety of pump powers injected at 11.961 GHz into the cavity. From this measurement, it is clear that the bandwidth, $BW$, of third harmonic generation increases with input pump power (up to the order of kHz) and is much greater than the linewidth of the high-$Q$ WG mode resonance (on the order of 10 Hz). This phenomena is typical for a nonlinear process. Figure \ref{gbwfig} compares, for the third harmonic, the product of the $SNR$ and $BW$ for a variety of frequencies corresponding to high $Q$-factor modes. From this data we see clear evidence of the existence of a $\chi^{(3)}$ nonlinearity due to the presence of Cr$^{3+}$ and Yb$^{+}$ impurity ions in sapphire, in addition to Fe$^{3+}$. The same modes were measured to determine if second harmonic generation occurs, however a much lower $SNR$ was observed and could not be considered significant over the values of $P_{res}$.

In summary, we show that a recently discovered nonlinear four-wave mixing effect due to Fe$^{3+}$ paramagnetic ions in sapphire can be used to create a frequency comb as well as generate third harmonic frequencies at a significant level. We have also shown that other species of paramagnetic ions (Cr$^{3+}$ and Yb$^{+}$) also exhibit a $\chi^{(3)}$ nonlinearity by measuring significant third harmonic generation by pumping at their ESR frequencies. The high frequency stability of frequency conversion using such nonlinearities was confirmed by determining the stability of the comb repetition rate exceeded that of a commercial Hydrogen maser over its bandwidth of $\sim40$ MHz. 
\begin{acknowledgments}
This work was funded by Australian Research Council grant numbers FL0992016 and CE11E0082.
\end{acknowledgments}

\bibliography{biblio}

\end{document}